\documentclass[a4paper,twocolumn,english]{revtex4}
\usepackage{graphicx}
\usepackage{amssymb}

\makeatletter

\providecommand{\LyX}{L\kern-.1667em\lower.25em\hbox{Y}\kern-.125emX\@}

\usepackage{babel}
\makeatother
\begin{document}

\title{Extended molecules and geometric scattering resonances in optical 
lattices}


\author{P.O. Fedichev,$^{1,2)}$ M.J. Bijlsma$^{1)}$, and P. Zoller$^{1)}$}

\address{$^{1}$ Institute for Theoretical Physics, University of Innsbruck,
A--6020 Innsbruck, Austria\\
 $^{2}$ RRC Kurchatov Institute, Kurchatov Square, 123456, Moscow,
Russia }

\begin{abstract}
We develop a theory describing neutral atoms scattering at low energies
in an optical lattice. We show that for a repulsive interaction, as
the microscopic scattering length increases, the effective scattering
amplitude approaches a limiting value which depends only on the lattice
parameters. In the case of attractive interaction a geometric resonance
occurs before reaching this limit. Close to the resonance, the effective
interaction becomes repulsive and supports a weakly bound state, which
can extend over several lattice sites.
\end{abstract}
\maketitle

Recent advances in cooling and trapping of cold atomic gases (see
\cite{BEC:review2003} for a review) have led to the experimental
realization of optically trapped lattices gases
\cite{mott:expBloch,Jaksch}. For a dilute gas and at low energies
all physical properties of such gases can be expressed in terms of
the two-body scattering amplitude \cite{LL:volIX}. It is well
established that confinement of a gas in one or more spatial
dimensions can strongly modify the collisional properties of
atoms, and can even induce geometric resonances
\cite{Olshanii1998a,petrov:1Dinteraction,petrov:2dscatter,petrov:confinement}.
In an optical lattice related phenomenona should occur when, (i)
the microscopic free-space scattering length becomes comparable to
the size of a single lattice well, a situation readily obtained by
combining Feshbach or shape resonances with optical trapping
techniques
\cite{Julienne1,feshbach:expjin,feshbach:expKetterle,feshbach:expmattW,feshbach:expSalomon};
and (ii) the tunneling between the neighboring lattice sites is
sufficiently small. Below we give a theoretical analysis of
the modification of neutral atoms scattering at low energies in an
optical lattice. This serves as the basis of writing down an
effective (long wavelength) field theory of a dilute optical
lattice gas with an effective atomic mass and effective atomic
scattering length, modified and tunable by the lattice parameters.

Let us consider neutral atoms confined in a 3D optical lattice and
interacting through a short range potential. Our goal is the
derivation of the low energy limit of the effective scattering
amplitude, and in particular the effective scattering length
$a_{\mathrm{eff}}$ in a tight binding model (TBM) as a function of
the free space scattering length $a$ and lattice parameters,
adapting techniques developed in
\cite{petrov:2dscatter,petrov:confinement}. We will show that the
wavefunction $\psi _{p}$ of the pair of atoms averaged over many
lattice sites satisfies the effective Schrdinger equation for
the relative motion,
\begin{equation} \label{SE2atoms}
\left(-\frac{\hbar
^{2}\nabla ^{2}}{m_{s}}+\frac{4\pi \hbar
^{2}}{m_{s}}a_{\mathrm{eff}}\delta (r)\frac{\partial }{\partial
|r|}|r|\right)\psi _{p}(r)=\epsilon _{p}\psi _{p}(r),
\end{equation}
where $\epsilon _{p}$ is the relative energy, and $m_{s}$ is the
band mass. The main results are: (i) For repulsive interaction,
when the microscopic scattering length $a$ increases, the
effective scattering amplitude remains finite and continuously
approaches a universal limit, which only depends on the optical
lattice parameters. (ii) In the case of an attractive interaction
a geometric resonance in the effective scattering amplitude,
similar to the resonances found in
\cite{petrov:2dscatter,petrov:confinement}, occurs before the
universal limit is reached. In addition, after crossing the
resonance the effective interaction becomes repulsive and weakly
bound states result close to the resonance. This predicts the
existence of large lattice induced molecules which can extend over
several lattice sites, which - by analogy with the Wannier-Mott
excitons in semiconductors \cite{semicond:YuCardona} - will be
referred to below as Wannier-Mott molecules. The parameters
effective mass $m_s$ and scattering length $a_{\mathrm{eff}}$
appearing in (\ref{SE2atoms}) are the relevant parameters for the
long wavelength effective field theory describing the dilute
lattice gas.

We turn now to a discussion of two-body scattering in an optical
lattice. The particles interact via a short range interatomic
potential $U(r)$ depending only on the relative coordinate $r$.
The range $r_{0}$ of this potential is assumed to be small
compared to the lattice spacing and to the size $l_{0}$ of the
ground state wavefunction in an individual well. In addition, the
lattice spacing is assumed to be small relative to the de Broglie
wavelength of the incoming particle. The particles move in an 3D
optical lattice created by counterpropagating laser beams. The
resulting periodic potential is characterized by a depth $V_{0}$
and lattice spacing $d$, which are assumed to be equal for all
spatial direction. Close to these minima, the potential can
expanded as
\begin{equation}
V=\frac{M\omega ^{2}}{2}\delta R^{2}+\frac{\mu \omega
^{2}}{2}\delta r^{2},\label{eq:reducedHam}
\end{equation}
where $\delta R$ and $\delta r$ are the deviation of the
center-of-mass and the relative coordinates $R=(x_{1}+x_{2})/2$
and $r=x_{2}-x_{1}$ from their potential minimum positions. Here,
$x_{1}$ and $x_{2}$ are the positions of the particles and $\mu
=m/2$ and $M=2m$ are the reduced and the total mass, respectively.
Moreover, the curvature within a single lattice site is given by
$\omega ^{2}=V_{0}/md^{2}$. From here on we will use units such
that $\hbar =m=1$.

In such a periodic potential, the two-body wavefunction $\Psi
(x_{1},x_{2})$ is characterized by certain quasimomenta $Q$ and
$q$ of the center of mass and the relative motion, as well as by
band indices $S$ and $s$. Using the fact that the lattice
potential is approximately separable (compare
Eq.~(\ref{eq:reducedHam})),  we can expand the two-body
wavefunction in a set of basis functions
\cite{LL:volIX,semicond:YuCardona},
\begin{equation}
\Psi_{Qq}^{(Ss)}=
\sum_{N,n}e^{i(Q,R_{N})+i(q,r_{n})}W_{S}(R-R_{N})w_{s}(r-r_{n}),\label{eq:twobodywavefunc}
\end{equation}
where $W$ and $w$ are Wannier functions. Within the TBM the
functions $W$ and $w$ coincide with the stationary states of the
Hamiltonian (\ref{eq:reducedHam}).

The wavefunctions of the relative motion (Bloch waves) are characterized
by the quasimomentum $q$, the band index $s$ and the energy
%
 $ \epsilon _{sq}=\epsilon _{s}-t_{s}\cos (qd)$,
%
where $t_{s}=\sqrt{D_{s}}\omega /\pi $, and $D_{s}\ll 1$ is the
WKB tunneling exponent between the neighboring wells, and $\epsilon _{s}$
are the energies of states $s$ in isolated wells \cite{LL:volIX}.
For sufficiently small momenta the dispersion relation can be approximated
as $\epsilon _{sq}\approx \epsilon _{s}+q^{2}/m_{s}$, where $m_{s}=2/t_{s}d^{2}$
is the effective mass. In the TBM, $l_{0}\alt d$, the tunneling coefficient
is small and hence $m_{s}\gg 1$. The asymptotic form of the relative
part of the two-body wavefunction far from the region around $r=0$
where the interaction potential acts, can be written as\[
\Psi (r)=\sum _{n}w_{s}(r-r_{n})e^{i(p,r_{n})}+\]
\begin{equation}
-\sum _{s^{\prime }n}\frac{m_{s}f_{ss^{\prime }}}{4\pi r_{n}}e^{iq_{s^{\prime }}r_{n}}w_{s^{\prime }}(r-r_{n}),\label{eq:scatteringdef}\end{equation}
where $f_{ss^{\prime }}$ is the $s-$wave scattering amplitude, and
$q_{s^{\prime }}=\sqrt{m_{s^{\prime }}(\epsilon _{sp}-\epsilon _{s^{\prime }0})}$.
This definition ensures that in the continuum case $f=4\pi a/m_{s}$,
where $a$ is the free space two-body scattering length and $m_{s}=m$.

Since the interparticle interaction $U(r)$ is very short-range, we
can use free (Bloch) solutions for the relative motion everywhere
apart from the immediate vicinity of $r=0$. To calculate the effective
scattering amplitude we closely follow the method suggested in \cite{petrov:2dscatter,petrov:confinement}.
The most general solution of the free Schroedinger equation for $r\neq 0$
for a given energy $\epsilon $, which also contains the incoming
wave is \begin{equation}
\Psi (r)=\sum _{n}w_{s}(r-r_{n})e^{i(p,r_{n})}+AG(r,0),\label{eq:oursolution}\end{equation}
where $A$ is an arbitrary constant. Here the first term describes
the incoming wave, whereas the second term gives the scattered waves.
The Green function is given by its usual expression\begin{equation}
G(r,r^{\prime })=\sum _{s^{\prime },q,n,n^{\prime }}\frac{w_{s^{\prime }}(r-r_{n})w_{s^{\prime }}(r^{\prime }-r_{n^{\prime }})}{\epsilon _{sp}-\epsilon _{s^{\prime }q}+i0}e^{i(q,r_{n}-r_{n^{\prime }})}.\label{eq:greenfuncdef}\end{equation}
For sufficiently large $r$ we can expand $\epsilon _{sq}$ up to
second order around $q=0$ and integrate over $q$ \begin{equation}
G(r,0)=-\frac{m_{s}V}{4\pi }\sum _{s^{\prime },n}\frac{\exp i(q_{s^{\prime }},r_{n})}{r_{n}}w_{s^{\prime }}(r-r_{n})w_{s^{\prime }}(0).\label{eq:scatteringamplitudedef}\end{equation}
where $V$ is the volume of the system. If $\epsilon _{s^{\prime }0}>\epsilon _{sp}$
then the corresponding partial wave does not propagate to $r\rightarrow \infty $,
$q_{s^{\prime }}$ is imaginary, and the channel $s^{\prime }$ is
closed. Comparing Eqs. (\ref{eq:scatteringdef}),(\ref{eq:oursolution})
and (\ref{eq:scatteringamplitudedef}) we identify the scattering
amplitude\[
f_{ss^{\prime }}(p,q^{\prime })=A(\epsilon _{sp})Vw_{s}(0)\delta _{ss^{\prime }},\]

The coefficient $A$ can be found by considering the short distance
assymptotics of the solution (\ref{eq:oursolution}). Indeed, at small
distances $r\ll l_{0}$ the wavefunction of the relative motion should
match the solution of the two-body scattering problem in 3D \cite{LL:volIII}:
\begin{equation}
\Psi (r)\rightarrow B(1-\frac{a}{r}),\label{eq:zerorangeboundary}
\end{equation}
where $B$ is a constant. The scattering length $a$ takes into account
all the processes occurring at distances of order $r_{0}\ll l_{0},d$.
In particular $a$ may contain the effect of Feshbach \cite{resonantscattering:holland}
or shape resonances \cite{shaperes:deutsch}, provided that the size
of the molecular bound state is sufficiently small.

We consider incoming particles at low energies so that the scattering
occurs in the lowest Bloch band $s=0$. Taking the limit both $r,r^{\prime }\rightarrow 0$
in the Eq.(\ref{eq:greenfuncdef}) we find\begin{equation}
G(r,r^{\prime }\rightarrow 0)=\sum _{s,q}\frac{w_{s}(r)w_{s}(r^{\prime })}{\epsilon _{op}-\epsilon _{sq}+i0}.\label{eq:greenfuncclosepoints}\end{equation}
As it is well known the Green function is singular when $r^{\prime }\rightarrow r$
and can be represented as \cite{chaos:gutzwiller}\begin{equation}
G(r,r^{\prime }\rightarrow 0)\approx -\frac{1}{4\pi |r-r^{\prime }|}+{\mathcal{F}}(\epsilon ),\label{eq:greenfuncshortdistance}\end{equation}
where ${\mathcal{F}}$ is the regular part of the Green function.
The singular part comes from the direct classical trajectory connecting
the points $r,r^{\prime }$, which has a universal character. The
regular part consists of the contribution of recurrent trajectories
and depends strongly on the details of the confining potential.

The only resonance in the denominator of Eq.(\ref{eq:greenfuncclosepoints})
occurs for the $s=0$ contribution. We should therefore study this
term separately by writing $G=G_{0}+G^{\prime }$, where the function
$G_{0}$ is not singular, and
\begin{equation}
G_{0}=\sum _{q}\frac{w_{0}(0)w_{0}(0)}{\epsilon _{0p}-\epsilon _{0q}+i0}.\label{eq:GFzerocomp}
\end{equation}
The TBM wavefunctions can be represented as $w_{s}=Z_{s}\psi _{s}(r)$
where $\psi _{s}$ is the wave function of the oscillator in a state
$s$, and $Z_{s}$ is a normalization factor ($Z_{s}^{2}=N^{-1}$,
where $N$ is the number of the lattice sites). Substituting the wavefunctions
in Eq.(\ref{eq:GFzerocomp}) we find for the imaginary part of $G_{0}$:
\begin{equation}
{\mathrm{Im}}\;G_{0}(p)=-i\frac{m_{s}}{4\pi }d^{3}|\psi
_{0}(0)|^{2}p.\label{eq:FRe}
\end{equation}
The real part is given by the principle value
\begin{equation} {\rm
Re}\;G_{0}(p\rightarrow 0)=-\frac{2\ln 2|\psi _{0}(0)|^{2}}{\pi
t_{0}}.\label{eq:FIm}
\end{equation}

The remaining terms in Eq.(\ref{eq:greenfuncclosepoints}) can be
analyzed in the WKB approximation. In a spherically symmetric harmonic
trap the spectrum of the excitations is $\epsilon _{s}=\omega (n_{x}+n_{y}+n_{z}+3/2).$
The WKB wavefunctions in each direction look like\[
w_{n}(x)=(\frac{\omega }{\pi })^{1/2}\frac{\cos (\Phi _{n}(x))}{(4\epsilon _{n}-\omega ^{2}x^{2})^{1/4}},\]
where $\Phi _{n}(x\rightarrow 0)\approx \pi n/2+qx$ with $q=\sqrt{\omega (n+1/2)}$.
Then\[
G^{\prime }=-\frac{\omega ^{1/2}}{8\pi ^{3}}\sum _{n_{x},n_{y},n_{z}}\frac{\Pi _{i=x,y,z}\cos (\Phi _{n_{i}}(x))\cos (\Phi _{n_{i}}(x^{\prime }))}{\sqrt{\Pi _{i=x,y,z}(n_{i}+1/2)}(n_{x}+n_{y}+n_{z})},\]
where the summation occurs over even values of $n$. Note that the
term with $n_{x}=n_{y}=n_{z}=0$ is absorbed in $G_{0}$ and hence
is excluded from the sum. This sum diverges when $x=x^{\prime }$.
Changing from summation to integration and separating out the singular
contribution we find:\[
G^{\prime }(r\rightarrow 0,0)=-\frac{1}{4\pi r}+\sum _{p}(\frac{1}{p^{2}}-\frac{1}{p^{2}-3\omega /2}).\]
The sum over $p$ can be estimated to give a contribution $\sim 1/l_{0}$
to ${\mathcal{F}}$. Within the TBM ($D_{0}\ll 1$) this is much smaller
than the contribution given by ${\mathrm{R}e}G_{0}$ and can therefore
be neglected:
\[
{\mathcal{F}}=-\frac{2\ln 2}{\pi tl_{0}^{3}}-i\frac{m_{s}d^{3}}{4\pi l_{0}^{3}}p.
\]

\begin{figure}
\includegraphics[  scale=0.40,angle=270,origin=lB]{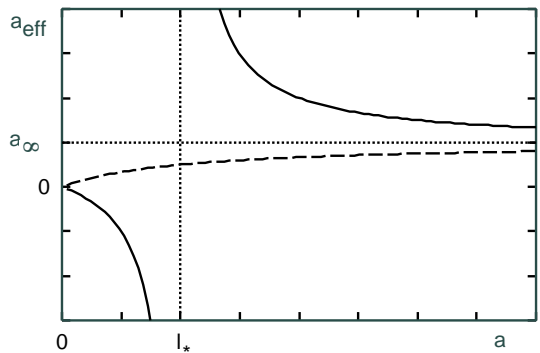}
\caption{ Dependence of the effective scattering length
$a_{\mathrm{eff}}$ on the absolute value of the microscopic
scattering length $|a|$. The dashed line corresponds to repulsive
interaction potential ($a>0$). The solid line corresponds to the
attractive potential ($a<0$) with a geometric resonance at
$|a|=l_{*}$.\label{effectivescattering}}
\end{figure}
Substituting Eq.~(\ref{eq:greenfuncclosepoints}) in
Eq.~(\ref{eq:oursolution}) and comparing the result with the
boundary condition (\ref{eq:zerorangeboundary}) we can compute the
coefficients $A$ and $B$ to find the expression for the scattering
amplitude:
\begin{equation} f_{00}(p)=\frac{4\pi
/m_{s}}{a_{\mathrm{eff}}^{-1}+ip},\label{eq:scamplitude}
\end{equation}
where $\beta =(d/l_{0})^{3}$, and the effective scattering length
is defined as
\begin{equation} \label{aeff}
 a_{\mathrm{eff}}=\frac{a\beta m_{s}}{1+a/l_{*}},
\end{equation}
with $l_{*}=l_{0}D_{0}^{1/2}/4\ln 2$.

 The dependence of the effective scattering
length $a_{\mathrm{eff}}$ on the microscopic scattering length $a$
is depicted in Fig.~1. First, according to Eq.~({\ref{aeff}}) in
the limit of small $a<<l_{*}$, we have $a_{\mathrm{eff}}/a = \beta
m_s
>>1$, i.e. the effective interaction in the lattice in comparison
with free space is strongly enhanced by both the strong
confinement of atoms in the lattice sites $\beta\equiv
(d/l_0)^3>>1$ and the large effective mass $m_s$. In the case of
repulsive interaction, $a>0$, $a_{\mathrm{eff}}$ continuously
grows as $a$ increases and approaches the universal value
$a_{\infty }=m_{s}\beta l_{*}\sim d\agt l_{0}$. In the case of
attractive interaction, $a<0$, the effective scattering length
exhibits resonant behavior and diverges at $|a|=l_{*}$. This is
the same type of {\em geometric resonance} suggested for quasi-1D
\cite{petrov:1Dinteraction} and quasi-2D gases
\cite{petrov:2dscatter,petrov:confinement} at low energies. At
larger values of $|a|$ the effective scattering becomes repulsive
and reaches the universal limit $a_{\infty }$.

The imaginary part in Eq.(\ref{eq:scamplitude}) is required by the
optical theorem \cite{LL:volIII} and does not imply any inelastic
process (apart from those possibly contained in ${\mathrm{Im}}a$).
Depending on the sign $a_{\mathrm{eff}}$ the poles of the
scattering amplitude give rise either to bound or to virtual
states in the effective potential as a result of the interplay of
the interparticle interaction and the lattice potential. An
interesting possibility arises when the microscopic scattering
length $a$ is negative and its absolute value is slightly larger
than the resonance value $l_{*}$. In this case the effective
interaction is repulsive and is characterized by a very large and
positive scattering length
\[
a_{\mathrm{eff}}^{+}=\frac{l_{*}\beta m_{s}}{|a|-l_{*}}.
\]
Accordingly, the scattering amplitude (\ref{eq:scamplitude}) has
a pole at $|p|=p_{*}$ corresponding to a bound state with a binding
energy
\[
|\epsilon _{*}|=\frac{p_{*}^{2}}{m_{s}}=\frac{1}{m_{s}(a_{\mathrm{eff}}^{+})^{2}}.
\]
The size of such a dimer is $\xi \sim a_{\mathrm{eff}}^{+}$ and
must be very large: $\xi \gg d$ so that the bound state of the two
particles can extend over many lattice sites, and thus should be
referred to as ``Wannier-Mott'' molecules in analogy to the
Wannier-Mott excitons in semiconductors \cite{semicond:YuCardona},
where bound states of electron and holes can very large compared
with the lattice constant.

Note that in the TBM $l_{*}\ll l_{0}$ and, therefore, the
geometric resonance $|a|\sim l_{*}$ can occur at realistic values
of the microscopic scattering length $a$. Because the rate of
3-body recombination scales as $a_{\mathrm{eff}}^{4}$
\cite{fedichev:3body} the spontaneous formation of these molecules
should be facilitated in the limit that the effective scattering
length becomes large. However, at the same time, in this limit the
binding energy is very small and a reasonable fraction of bound
atoms can only be expected at temperatures $T\ll |\epsilon _{*}|$.
A possible way of detecting these extended and weakly bound
Wannier-Mott molecules arises if they are composed of fermionic
atoms in different internal states. When the trapping potentials
for the separate states are tilted in opposite directions, the
unbound atoms experience a force and will move in one or the other
direction, depending on their internal state, whereas the
molecules experience zero net force. If the potential gradient
difference exceeds a certain critical value related to the binding
energy, the molecules dissociate and the remaining atoms are
released.

The limiting value of the 2-body scattering amplitude $f_{00}(0)$
is of particular importance for condensed matter calculations\[
f_{00}(0)=\frac{4\pi a_{\mathrm{eff}}}{m_{s}}=\frac{4\pi \beta }{a^{-1}+l_{*}^{-1}}.\]
 and determines the effective interaction strength $\tilde{U}\equiv f_{00}(0)$
for a dilute lattice gas. For example, in the case of a Bose-condensed
lattice gas the chemical potential is given by $\tilde{U}n$, where
$n$ is the density of the condensed particles \cite{Beliaev}. Similarly,
in a system of interacting fermions with an attractive interaction
($\tilde{U}<0$) the critical temperature of the BCS transition is
\cite{BCS:GorkovMelikborh} \begin{equation}
T_{c}=\frac{8\exp (\gamma -2)}{\pi }\epsilon _{F}\exp (-2\pi ^{2}/p_{F}\tilde{|U}|),\label{eq:TCdef}\end{equation}
where $\epsilon _{F}=p_{F}^{2}/2m_{s}$ is the Fermi energy. An increase
of the absolute value of the scattering length leads to a resonance
in the effective interaction potential and causes the latter to change
from an attractive to a repulsive interaction at $|a|=l_{*}$. Note
however, that close to the resonance the imaginary and energy dependent
part of the scattering amplitude cannot be neglected and Eq.(\ref{eq:TCdef})
loses its validity. A further increase of $|a|$ leads to a repulsive
interaction which destroys the BCS ground state and results in the
formation of dimers. Since the latter are bosons, at sufficiently
small temperatures a BEC of dimers can be created. This possibility
is related to the ongoing discussion of the BCS-to-BEC crossover in
strongly interacting Fermi-gases with attractive interaction \cite{BCSBEC:nozieres}.
We note that the BEC of large molecules provides an example of a superfluid
ground state for a system of interacting fermions with effective repulsion.
This possibility is related to the reported BEC of excitons in semiconductors
\cite{Lin1993a}.

In summary, we have derived an analytical expression for the
scattering amplitude as a function of the free space scattering
length and the parameters of the confining optical potential for
small energies. For a repulsive interaction, as the microscopic
scattering length increases, the effective scattering amplitude
grows and continuously approaches a universal limit. In the case
of attractive interaction, increasing the scattering length leads
to a geometric resonance occurring before the universal limit is
reached. Close to the resonance, when the effective interaction
becomes repulsive, it supports a weakly bound Wannier-Mott
molecule which can extend over several lattice sites.

Discussions with G.V. Shlyapnikov, A. Daley, A. Recati, U. R. Fischer are
gratefully acknowledged. Work at Innsbruck supported in part by
the Austrian Science Foundation, EU Networks, and the Institute
for Quantum Information. P.F. thanks RFBR for support.

\end{document}